\documentclass[preprint,12pt]{elsarticle}

\usepackage{amssymb}
\usepackage{amsmath}

\usepackage{lineno}
\usepackage{url}
\usepackage{todonotes}
\usepackage{layout}
\usepackage{booktabs}
\usepackage{makecell}
\usepackage{multirow}
\usepackage{doi}

\usepackage{printlen}

\journal{Computers \& Fluids}

\begin{document}

\begin{frontmatter}



\title{Bathymetry Reconstruction by Bayesian Inference}


\author[tuhh]{Lars Stietz\fnref{uhh}\corref{cor1}} 
\ead{lars.stietz@tuhh.de}
\author[tuhh]{Sebastian G\"otschel}
\author[uhh]{Peter Schleper}
\author[tuhh]{Daniel Ruprecht}

\affiliation[tuhh]{organization={Chair Computational Mathematics, Institute of Mathematics, Hamburg University of Technology},
            addressline={Am Schwarzenberg-Campus 3},
            city={Hamburg},
            postcode={21073},
            country={Germany}}
\affiliation[uhh]{organization={Institute of Experimental Physics, University of Hamburg},
            addressline={Luruper Chaussee 149},
            city={Hamburg},
            postcode={22761},
            country={Germany}}
\cortext[cor1]{Corresponding author: Lars Stietz}

\begin{abstract}
Bathymetry reconstruction is an important problem in various fields, including oceanography and environmental monitoring.
This paper presents a Bayesian inference approach to reconstructing bathymetries from point measurements of the water height.
We test the method for parameterized and discretized bathymetries with synthetic data to evaluate its performance and limitations.
Our results indicate that the Bayesian framework provides a robust approach to bathymetry reconstruction.
Finally, we use the framework to reconstruct a real-world bathymetry in a wave flume from experimental measurements and compare its performance to an adjoint optimization method.
The Bayesian approach improves the normalized root mean squared error (NRMSE) of the reconstruction and provides better qualitative features, while also quantifying uncertainty.
\end{abstract}


\begin{highlights}
\item We propose an approach to bathymetry reconstruction from sparse data based on Bayesian inference
\item Quantification of uncertainties is achieved by using Markov chain Monte Carlo (MCMC) methods
\item We demonstrate efficacy of the approach for a parameterized low dimensional and discretized high dimensional bathy\-metry and synthetic as well as experimental data
\end{highlights}


\end{frontmatter}

\hyphenation{ba-thy-me-try}



\section{Introduction}\label{sec:introduction}
Knowledge of the bathymetry and therefore water depths of oceans, lakes, and rivers, is crucial for various scientific and engineering applications.
Accurate bathymetry information is essential for example for tsunami modeling and forecasting \cite{behrens2010new}, rip current prediction \cite{dalrymple2011rip} and flood risk assessment \cite{neal2021estimating}.
A systematic review of remote sensing methods for bathymetry reconstruction discusses a range of air- and space-borne measurements using various techniques~\cite{heRemoteSensingShallow2024}.
Methods reconstructing the bathymetry from LiDAR measurements, imagery \cite{collinsBathymetricInversionUncertainty2020,oadesEvaluationNearshoreBathymetric2023}, or radar are commonly used, but also in-situ measurements using water wave properties have been shown to be effective \cite{vasanInverseWaterWave2013,angel24bathy}.
If the whole surface wave field is known, the bathymetry can be reconstructed by direct inversion of the shallow water equations~\cite{noureddineDirectApproachDetection2025}.
However, in many situations only sparse measurements of the free surface elevation are feasible and direct measurements of the bathymetry are expensive and difficult.

Inverse problems, where unknown parameters or system states need to be inferred from observed data, arise in many scientific domains.
They are often ill-posed with non-unique solutions and sensitivity to noise.
Multiple techniques have been developed to tackle these challenges, including regularization methods and statistical Bayesian approaches~\cite{kaipioStatisticalComputationalInverse2005}.
Methods to infer fluid properties or unknown flow conditions, like the bathymetry, have been considered in the framework of inverse problems~\cite{sellierInverseProblemsFree2016}.

Formulating bathymetry reconstruction as an inverse problem taking a Bayesian perspective allows for a more comprehensive understanding of the problem, including the ability to quantify the uncertainty of the reconstruction results.
Both the regularization and Bayesian approach can incorporate prior knowledge about the system and unknown parameters to  restrict the solution space and mitigate the ill-posedness of the problem.
Comprehensive mathematical foundations exist for the conversion of the classical to Bayesian inverse problems~\cite{kaipioStatisticalComputationalInverse2005,stuartInverseProblemsBayesian2010a}.

In this work, we investigate the application of Bayesian inference to bathymetry reconstruction and compare its performance to an adjoint-based approach using very sparse measurements~\cite{angel24bathy}.
Ruppenthal et al. \cite{ruppenthal2026bathymetry} also use a PDE-constrained approach and show the effectiveness of $L_1$ - regularization to suppress noise, but resolve the whole spatial domain.
Wu et al. \cite{wuAdjointbasedHighorderSpectral2023a} also use adjoint-based optimization with a high-order spectral method and a velocity potential satisfying the Laplace equation as forward model.
Both are applied on simulated data only.
A Bayesian approach for a similar problem, reconstructing the initial state for the Tohoku tsunami for a given bathymetry,  using a multi layer Markov chain Monte Carlo (MLMCMC) method is presented by Seelinger et al.~\cite{SeelingerEtAl2021}.
Bayesian inversion using Markov chain Monte Carlo (MCMC) was tested for the reconstruction of near shore bathymetries from stationary surface wave properties using a linear forward model~\cite{lasithadhikariInvertingShoreBathymetry2016}.
By contrast, the forward model we consider in this work is nonlinear and time-dependent and the reported NRMSE of  50.35\% for simulated data and around 29.25\% for real-world data are higher than the 10\% we report by by using time-resolved information.
Near shore bathymetry reconstruction including uncertainty estimation on surf-zone imagery has been tested using a machine learning approach in \cite{collinsBathymetricInversionUncertainty2020}.
They show robustness for large wave heights, but only for synthetic data.
Another machine learning approach including Bayesian inversion for bathymetry reconstruction is presented by Ramesh et al.~\cite{rameshGATSBIGENERATIVEADVERSARIAL2022} as a proof of concept, but only tested on simulated data.

This paper makes the following contributions.
\begin{enumerate}
    \item We formulate the problem of bathymetry reconstruction from time-dependent data from the perspective of Bayesian inference.
    \item We show that the resulting Bayesian inverse problem can be solved by using a Markov chain Monte Carlo (MCMC) method that imploys non-linear shallow water equations as forward model.
    \item We analyze the MCMC approach for a simple low dimensional parameterized bathymetry reconstruction test with simulated data and show that this can deliver insights into the structure of the inverse problem.
    \item We apply the method to real-world experimental data to reconstruct a non-parameterized discretized bathymetry and compare the results with the results obtained using an adjoint optimization method~\cite{angel24bathy}.
\end{enumerate}
Section \S\ref{sec:method} describes the Bayesian framework and the setup on which it will be tested.
Then, Section\S\ref{sec:results} tests the approach first for a parameterized and then a discretized bathymetry using synthetic data and finally for a discrete bathymetry using real-world data from a dedicated experimental setup.
Conclusions are drawn in \S\ref{sec:conclusion}.
To reproduce the results the used code is available at \cite{myCode} and the data can be found at \cite{dataset}.
\section{Methodology and Problem Description}\label{sec:method}
Inferring the bathymetry from time-resolved measurements of the free surface elevation $H$ is an inverse problem.
The reconstruction process requires a forward model that maps a given hypothesis about the bathymetry to the corresponding measurements.
Since we consider a channel geometry as experimental setup, we use the time-dependent one-dimensional shallow water equations (SWEs)

\begin{align}
    \begin{pmatrix}
        h\\
        u
    \end{pmatrix}_t +
    \begin{pmatrix}
        hu\\
        \frac{1}{2}u^2 + gh
    \end{pmatrix}_x =
    \begin{pmatrix}
        0\\
        -gb_x -\kappa u
    \end{pmatrix}\label{eq:swe}
\end{align}
which have been shown to be a computationally efficient and suitable model~\cite{angel24bathy}.
In~\eqref{eq:swe}, $h = H - b$ is the water height above ground, $u$ the velocity, $g$ the gravitational acceleration, $b$ the bathymetry and $\kappa$ a bottom-friction coefficient.
For our choice of parameters, boundary conditions, and initial conditions we follow Angel et al.~\cite{angel24bathy}, except that we use a time-step of $\Delta t= 10^{-2}$ to match the measurement frequency.
Tests not documented here show that the larger time-steps has no significant impact on the accuracy of the forward model whilst reducing computational cost.
We use the open-source spectral solver framework Dedalus~\cite{burns20dedalus} for the numerical solution of~\eqref{eq:swe}.

\subsection{Experimental Setup}\label{sec:experimental_setup}
\begin{figure}
    \includegraphics[width=\textwidth]{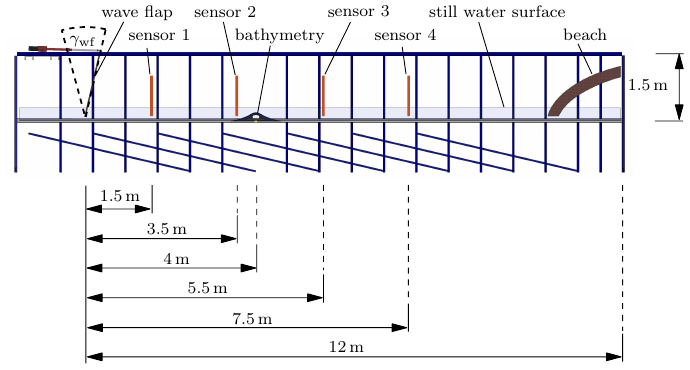}
    \caption{Sketch of the wave flume with sensor positions in orange and the bathymetry in black. From Angel et al. \cite{angel24bathy}, $\copyright$ 2024 The Author(s). Published by Elsevier Ltd. The generated data is publicly available~\cite{dataset}.}
    \label{fig:wave_flume}
\end{figure}

We evaluate our approach using experimental data from a wave flume experiment~\cite{dataset}.
In total, we have 20 independent repeated measurements with the same bathymetry and experimental setting.
To evaluate our method, we use the mean of the 20 measurements at each sensor as the observed measurements $H_\mathrm{obs}$ for inference.
Measurements of both synthetic and experimental data are taken at 4 equidistant points $x_\mathrm{sensor} \in \{1.5,3.5,5.5,7.5\}$ m at a frequency of $100$ Hz.
Figure~\ref{fig:wave_flume} shows a sketch of the experimental setup with sensor positions indicated by orange bars.
The first sensor is used as the left boundary condition for the forward model~\eqref{eq:forward_problem} while data from the three remaining sensors is used for reconstruction.
A bump manufactured from skateboard ramps is placed in the flume as bathymetry with its center positioned at 4 m and shown in black in Figure~\ref{fig:wave_flume}.
The bathymetry is flat before and after the bump.
To obtain a ground-truth for the discretized  bathymetry, we use measurements of the bathymetry~\cite{dataset} and interpolate the values to the mesh points of the discretization used for the forward model.
For the interpolation we use SciPy's PCHIP 1-D monotonic cubic interpolation similar to the one used in~\cite{angel24bathy}.
A more detailed description of the experiment was published together with the generated data~\cite{dataset}.

\subsection{Bayesian Inversion}
Let $F$ be the forward model that maps a bathymetry $b$ to approximate measurements $H:=h+b$ by solving~\eqref{eq:swe} numerically so that $ H \approx F(b)$.
The error introduced by noise in the measurements is modelled as additive Gaussian noise with zero mean and covariance $\Sigma$, so that
\begin{align}
     H = F(b) + \eta , \quad \eta \sim \mathcal{N}(0, \Sigma).\label{eq:forward_problem}
\end{align}
We assume independence of the noise in each sensor and over time so that the covariance matrix $\Sigma$ in \eqref{eq:forward_problem} is diagonal.

The inverse problem is to find the bathymetry $b$ from noisy measurements of $H$.
From a Bayesian inference perspective, we view the measurements and bathymetry as random variables.
Solving the inverse problem means finding the posterior distribution $\pi(b|H_\text{obs})$, that is to determine the probability that a bathymetry $b$ is correct given some measurements $H_\text{obs}$.
Using Bayes' theorem, the posterior distribution can be expressed as
\begin{align}
    \pi(b|H_\mathrm{obs}) = \frac{\pi_\text{L}(H_\text{obs}|b)\pi_\mathrm{Prior}(b)}{\pi(H_\mathrm{obs})} \propto \pi_\mathrm{L}(H_\mathrm{obs}|b)\pi_\mathrm{Prior}(b),
\end{align}
where $\pi_\mathrm{L}(H_\mathrm{obs}|b)$ is the likelihood function and $\pi_\mathrm{Prior}(b)$ the prior density for the bathymetry.
The likelihood function can be derived from the forward problem~\eqref{eq:forward_problem} and noise~\eqref{eq:likelihood_variance}~\cite{stuartInverseProblemsBayesian2010a,kaipioStatisticalInverseProblems2007}, leading to
\begin{align}
    \pi_\mathrm{L}(H_\mathrm{obs}|b) = \prod_{i=1}^{n_s}\prod_{t=1}^{T}\left(\frac{1}{\sqrt{2\pi\sigma_i^2}}\right)\exp\left(-\frac{1}{2\sigma_i^2}\|H_{\mathrm{obs},i}^{(t)} - F_i^{(t)}(b)\|_2^2\right) \label{eq:likelihood},
\end{align}
where $T \in \mathbb{N}$ is the number of timesteps and $n_s \in \mathbb{N}$ the number of sensors for a total of $d = T n_s$ data points.
We consider measurements from three independent sensors with a frequency of 100 Hz, leading to $d=3000$ measurements of $H$ in total.
To estimate the variance $\sigma_i^2$ of the measurement noise in~\eqref{eq:likelihood}, we use a simulated measurement of a flat bathymetry and compare it to measurements, so that
\begin{align}
    \sigma_i^2 =\frac{1}{T} \sum_{t=1}^{T}\left(H_{\mathrm{obs},i}^{(t)} - F_i^{(t)}(\mathbf{0})\right)^2\ \label{eq:likelihood_variance},
\end{align}
where $H_{\mathrm{obs},i}^{(t)}$ are the observed measurements from sensor $i$ at time step $t$, and $F_i^{(t)}(\mathbf{0})$ are the simulated measurements from the forward model with a flat bathymetry $b(x) = 0$.

The prior distribution $\pi_\mathrm{Prior}(b)$ encodes whatever knowledge or assumptions about the bathymetry are available in the given context.
In this work, we consider different prior distributions, particularly Gaussian and Cauchy distributions, to investigate the impact of the prior on the reconstruction results, see Sections~\ref{sec:parametrized_bathymetry_reconstruction} and \ref{sec:discretized_bathymetry_reconstruction}.

Since our forward model~\eqref{eq:swe} is nonlinear, it is not possible to derive an analytical expression for the posterior distribution.
Various techniques for exploring the posterior approximately exist~\cite{stuartInverseProblemsBayesian2010a}.
We use a Markov chain Monte Carlo (MCMC) method to sample from the posterior distribution and derive summary statistics, such as the mean, the standard error and a 95\% credible interval, to get an estimate of the bathymetry as well as the uncertainty of this estimate.
We evaluate the MCMC samples using the \texttt{MCMCChains.jl} package contained in the Julia \texttt{Turing.jl} library~\cite{pmlr-v84-ge18b, 10.1145/3711897}.
Since the scope of this work is to evaluate the potential of the Bayesian inference approach for bathymetry reconstruction, we leave further optimization of the sampling process for future work.

We use the Metropolis-Hastings (MH) algorithm~\cite{metropolisEquationStateCalculations1953, hastingsMonteCarloSampling1970} for sampling, because of its simplicity.
Tests comparing the MH algorithm with the preconditioned Crank-Nicolson algorithm (pCN)~ \cite{cotterMCMCMethodsFunctions2013} algorithm showed that the MH algorithm achieves similar accuracy in our case.
More advanced MCMC methods exist, such as Hamiltonian Monte Carlo~\cite{duaneHybridMonteCarlo1987}, No-U-Turn sampler~\cite{hoffmanNoUTurnSamplerAdaptively} or the Metropolis-adjusted Langevin algorithm~\cite{robertsExponentialConvergenceLangevin1996}, but require gradient information of the forward model.
Including gradient information in the sampling process is possible but would require a computationally costly backward evaluation of the forward model in every iteration, similar to adjoint optimization methods.
This is also left for future work.
The proposal distribution used in the MH algorithm is tuned to achieve an acceptance rate within the interval $[0.1, 0.4]$, which is in line with what is considered optimal in the literature~\cite{robertsOptimalScalingVarious2001a}.

We test the approach both for the low-dimensional case of a parameterized bathymetry and the high-dimensional case of a discretized bathymetry.
In the parameterized bathymetry test, we assume the shape of the bathymetry is a Gaussian bump and we only need to infer its width $b_w$ and position $b_p$.
In the discretized bathymetry test, we discretize the spatial domain using $n_x$ equidistant grid points and infer the height of the bathymetry at each grid point.
Note that we use different discretizations for the forward model and the bathymetry as the solver has different requirements on spatial resolution, in particular at the boundaries of the domain.
To calibrate the setup for the experimental data, we use simulated measurements of the free surface elevation $H_i$ at each sensor location generated by the forward model with a known bathymetry.

\section{Numerical Results}\label{sec:results}
This section evaluates the performance of Bayesian inference for reconstructing the bathymetry in the setup described above.
We first consider synthetic measurements where the forward model is used with a known bathymetry to generate data.
Gaussian noise with a standard deviation of 5\% of the maximum wave height is added to these synthetic measurements to simulate realistic noise~\cite{angel24bathy}.
Synthetic data is generated with a different discretization than in the inference to avoid an inverse crime \cite{kaipioStatisticalInverseProblems2007}.
For the synthetic data, we use 128 grid points and a time step of $\Delta t = 5\cdot 10^{-5}$ for the forward model, while the inference uses 64 grid points and a time step of $\Delta t = 10^{-2}$.

To explore the impact of the choice of prior, we first investigate the approach for a low-dimensional reconstruction of a Gaussian shaped bathymetry, where only the position and width need to be determined.
Then, the choice of priors is further refined for a high-dimensional setup by reconstructing a fully discrete bathymetry from synthetic data.
The best performing setup using both a sparse and smoothness prior is then applied to the experimental data set.
Achieved reconstruction errors are compared against the adjoint optimization approach~\cite{angel24bathy}.
We use the normalized root-mean-square error (NRMSE)
\begin{align}
    \label{eq:nrmse}
    \mathrm{NRMSE} = \frac{\sqrt{\frac{1}{N}\sum_{i=1}^{N}(b(x_i) - b_i)^2}}{\max(b(x)) - \min(b(x))},
\end{align}
the relative $\mathrm{L}_2$ and relative $\mathrm{L}_\infty$ error to quantify the error between the reconstructed bathymetry $b_i$ and the exact bathymetry $b(x_i)$ at the grid points $x_i$.
As reconstructed bathymetry, we use the mean of the MCMC samples at each grid point.
To obtain mean values at each grid point in case of the parameterized bathymetry, we evaluate each sampled Gaussian bathymetry at the grid points.

We use $N=64$ equidistant grid points for the reconstruction of the bathymetry and sampling step in the MCMC algorithm while the forward model uses a non-equidistant spectral discretization with the same number of points.
PCHIP interpolation from~\texttt{DataInterpolations.jl}~\cite{Bhagavan2024} is used to interpolate the sampled bathymetry values to the spectral grid in the SWEs solver.
If prior knowledge about the bathymetry is available, the sampling mesh could be refined in relevant areas, but this approach is not explored here.
In all tests, the likelihood model is given by equation~\eqref{eq:likelihood}.
For the MCMC, we use 2000 samples and a burn-in of 200 samples for the parameterized bathymetry and 5000 samples with a burn-in of 1000 samples for the discretized bathymetry.

\subsection{Reconstruction of a Parameterized Bathymetry}\label{sec:parametrized_bathymetry_reconstruction}
We assume that the bathy\-metry is a Gaussian
\begin{align}
	\label{eq:b_par}
    b(x; b_p, b_w) = 0.2 \exp\left(-\frac{(x-b_p)^2}{2b_w}\right).
\end{align}
defined by its center $b_p$ and width $b_w$ with a fixed and known height of $0.2$ m.
Least squares minimization, comparing the exact experimental bathymetry with the assumption from \eqref{eq:b_par}, provides the target parameters $b_p\approx 4.0$ and $b_w\approx 0.05$ for which~\eqref{eq:b_par} matches the real bathymetry most closely.
The small resulting NRMSE of 0.002 shows that the assumption of a Gaussian bathymetry is valid, but lacks in capturing the flat tails of the exact bathymetry.

We start with weak priors that only limit the position of the Gaussian to be inside the wave flume and away from the boundaries, that is ${b_p \sim \mathcal{U}([1.5,12.5])}$.
We limit the width to be between within ${b_w \sim \mathcal{U}([0, 1])}$.
For the sparse measurements of the wave height in this problem, we observe that Markov chains with different initial values in the $(b_p, b_w)$ space can converge to different local maxima of the posterior distribution.
To study the range of parameters for which the reconstruction is possible, we explore which priors sufficiently restrict the parameter space to guide the Markov chains towards the correct solution.

Testing different peak positions and widths, while keeping the other parameter fixed, reveals limitations of the reconstruction for bathymetries outside the sensor positions and for very narrow or wide peaks.
We create synthetic measurements with target peak positions in the interval $b_p \in [1.5,11.0]$ and fix $b_w=0.05$.
Analogously, we generate synthetic measurements where $b_w \in [0.01,0.5]$ while fixing $b_p=4$.
To evaluate the limits of the reconstruction of the peak position (peak width), we run the same MCMC settings for each peak position (peak width) measurement.
We run multiple MCMC chains with different initial values and discard chains that have an average log-posterior value more than 2 below the maximum average log-posterior value across all chains.

\begin{figure}
        \includegraphics{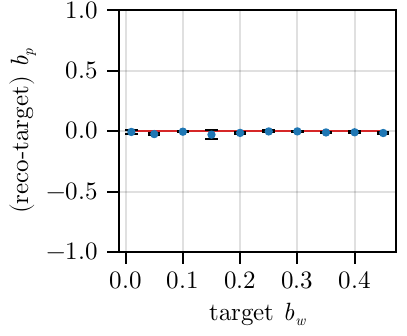}
        \includegraphics{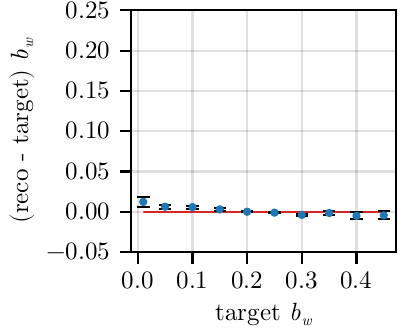}
    \caption{Difference between the reconstructed and simulated position (left) and width (right) of the parameterized bathymetry depending on its width.
        The red line shows a perfect reconstruction.
    Blue dots show the difference between the reconstructed and the target $b_w, b_p$.
    The error bars display the standard error of the MCMC samples to the reconstructed mean.}\label{fig:width_test}
\end{figure}

\begin{figure}
        \includegraphics{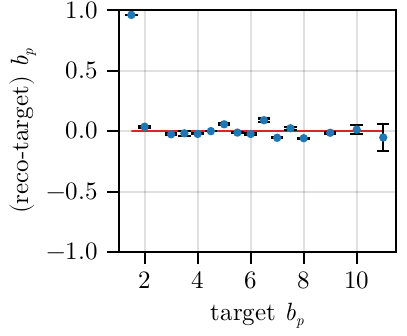}
        \includegraphics{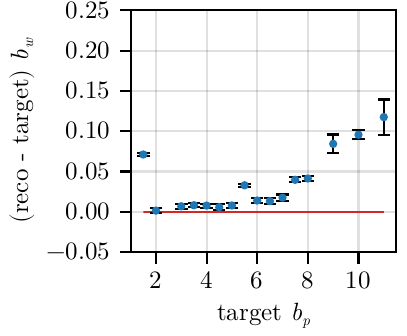}
    \caption{Difference between the reconstructed and simulated position (left) and width (right) of the parameterized bathymetry depending on its position.
    The red line shows a perfect reconstruction.
    Blue dots show the difference between the reconstructed and the true values of $b_w, b_p$.
    The error bars display the standard error of the MCMC samples to the reconstructed mean.
    While the position can be reconstructed reasonably accurately unless the bump is too close to the boundaries of the flume, the reconstruction of the width gets worse as the bump is placed further down the flume.
   }\label{fig:peak_test}
\end{figure}

In figures~\ref{fig:width_test} and~\ref{fig:peak_test}, we show the difference between the reconstructed (reco) and the target value of each parameter versus the target value of the varied parameter.
Additionally, we provide the standard error of the MCMC samples to the reconstructed mean as an estimate of the uncertainty of the reconstruction.
The standard error combines the estimated standard deviation of the MCMC samples with the number of effective samples, which is a measure of the number of independent samples in the MCMC chain after accounting for autocorrelation.

In Figure~\ref{fig:width_test} we keep the target $b_p=4.0$ fixed and vary the width $b_w$.
We can see that the position is reconstructed well for different widths in Figure~\ref{fig:width_test} (left).
The width is also reconstructed reasonably well in Figure~\ref{fig:width_test} (right), showing that the reconstruction overestimates the width for narrow peaks and underestimates the width for wide peaks with an increase in the uncertainty for both extremes.

Reconstructions of the peak shown in Figure~\ref{fig:peak_test} are more limited.
Here we fix the width to $b_w=0.05$ and vary the position $b_p$.
Consistent with the results of the width reconstruction we can see an overestimation of the peak width for the narrow peak of $b_w=0.05$.
The overestimation of the width increases with peaks positioned further away from the sensors, leading to a failure of the reconstruction for peaks positioned outside the interval $[2.0,7.5]$.
The reconstruction of the peak position only fails at the very end of the prior interval with a peak position below $2.0$.

\paragraph{Impact of the Choice of Prior}
Throughout the rest of this paper, synthetic measurements are understood to be generated from the exact experimental bathymetry.
We tested different prior distributions to restrict the parameters describing the bathymetry.
The weakest prior $b_p \sim \mathcal{U}([1.5, 12.5])$ simply restricts the position of the bump to within the spatial domain of the wave flume.
The width is restricted by the flat prior $b_w \sim \mathcal{U}([0, 1])$.
These uniform priors restrict the parameters to a reasonable range but assign equal probability to all values within, so they do not encode any additional knowledge about the bathymetry.
For illustration purposes, we use initial values of the MCMC chains of $b_w=0.5$ and $b_p \in \{2.5, 3.0, 4.0, 5.0, 5.5, 6.0\}$.

\begin{figure}[t]
    \includegraphics{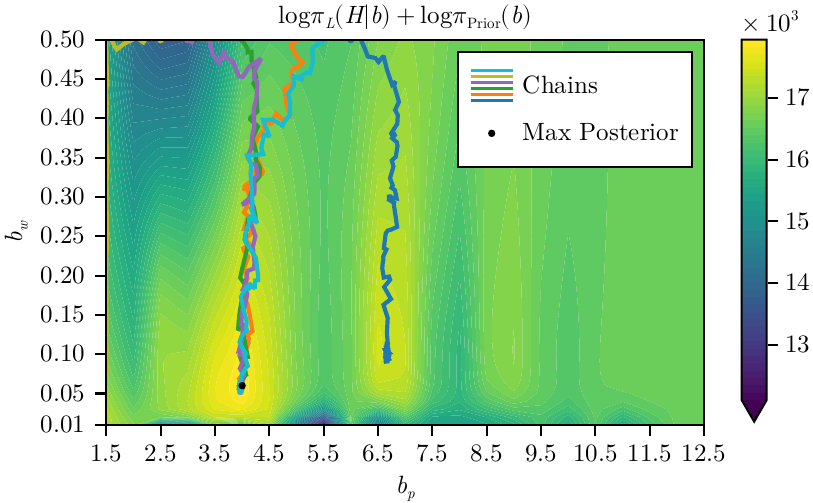}
    \caption{The non-normalized log-posterior distribution with uniform priors on the position and width.
    The lines show the MCMC chains with different initial values, where the initial values are marked with a dot.
    A black dot marks the global maximum of the posterior distribution.}
    	\label{fig:log_posterior_flat_prior}
\end{figure}

The resulting posterior landscape is shown in Figure~\ref{fig:log_posterior_flat_prior}.
For better visibility of the chains, we only show the part of the posterior landscape where $b_w\leq0.5$.
The posterior decays slowly for wider bathymetries, but the overall impact of the width is relatively weak, particularly for $0.5<b_w\leq1$.
It has multiple local maxima, illustrating the difficulties of reconstructing the correct bathymetry without prior information.
We can see that MCMC chains with different initial values converge to different local extrema, while the chain starting at $b_p=2.5$ diverges.

\begin{figure}[t]
    \includegraphics{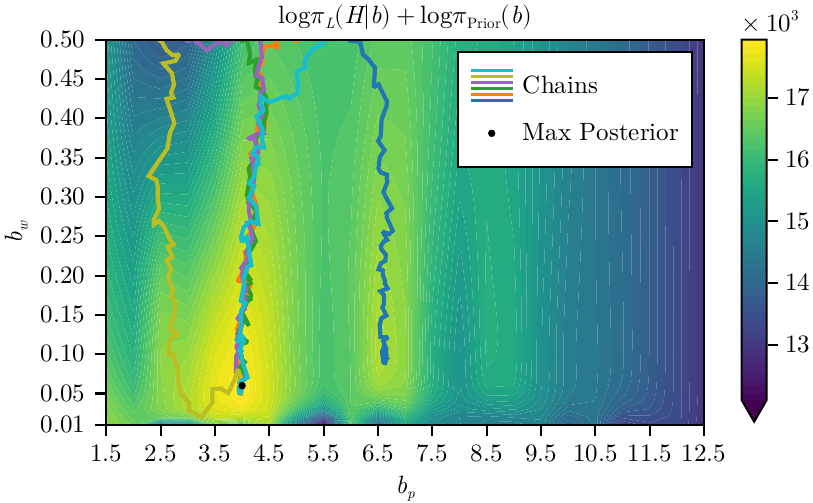}
    \caption{The non-normalized log-posterior distribution with prior on position ${b_p \sim \mathcal{N}(4,0.1)}$ and uninformative prior on width $b_w\sim \mathcal{U}([0,1])$.
    The lines show the MCMC chains with different initial values, where the initial values are marked with a dot.
    A black dot marks the global maximum of the posterior distribution.}
    \label{fig:log_posterior_gaussian_prior}
\end{figure}

To incorporate stronger knowledge about the bathymetry, we demonstrate the effect of a Gaussian prior for the position $b_p$ with a mean equal to the correct value and the variance controlling the uncertainty of our prior knowledge.
The resulting posterior landscape for $b_p \sim \mathcal{N}(4,0.1)$ is shown in Figure~\ref{fig:log_posterior_gaussian_prior}.
The Gaussian prior reduces the probability of the spurious local extrema for too small or too large values of $b_p$, which causes the MCMC chain starting at $b_p=2.5$ to now converge to the global maximum.
However, the MCMC chain starting at $b_p=6$ still gets stuck in a local maximum.
An even narrower Gaussian prior with a variance of $0.05$ would be required to get all chains to converge to the global maximum but having such a strong prior is unrealistic in practice.

\begin{figure}[t]
    \includegraphics{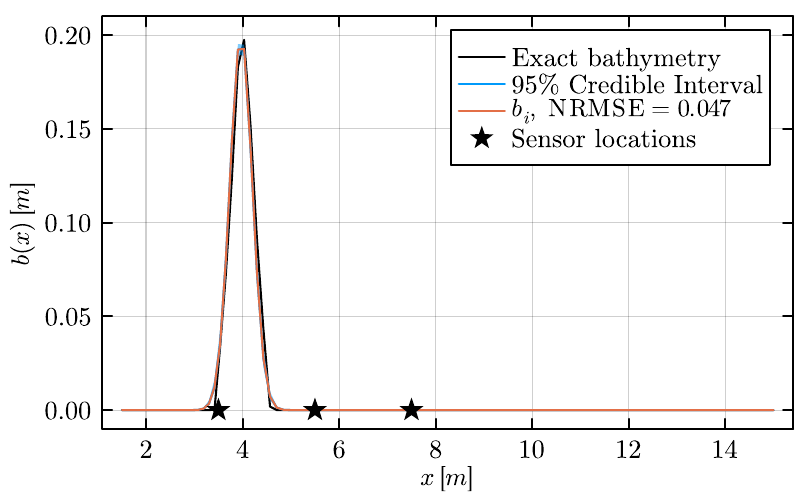}
    \caption{Sampled mean parameterized Gaussian bathymetry and 95\% credible interval reconstructed from experimental data from chain started at $b_p=5.0$ and $b_w=0.5$.
    We use a Gaussian prior for the position $b_p \sim \mathcal{N}(4,0.1)$ and a uniform prior for the width $b_w \sim \mathcal{U}([0, 1])$.
    The real bathymetry from the experiment is shown in black.
    Note that the difference in scaling of the x- and y-axis.
    The orange graph shows the mean of all sampled bathymetries from one chain at positions $x_i$ with a blue shaded region representing the 95\% credible interval.
    The shaded area is barely visible as the mean bathymetry is very close to the real bathymetry and the result is highly accurate.}
    \label{fig:parametrized_exp_result}
\end{figure}

To illustrate the reconstruction results, we show the sampled mean bathymetry and a 95\% credible interval for the chain starting at $b_p=5.0$ and $b_w=0.5$ in Figure~\ref{fig:parametrized_exp_result}.
This provides a low-dimensional baseline against which the reconstruction of the discretized bathymetry in \S\ref{subsec:experimental} can be compared.
With the used priors, we achieve a highly accurate reconstruction of the bathymetry with very high confidence.
Note that the low error of $0.029$ NRMSE reflects the fact that the Gaussian shape~\eqref{eq:b_par} is a good fit of the real bathymetry.
For a differently shaped bathymetry, the fitting error would eventually dominate the reconstruction error.
Using the same MCMC settings as before but with experimental measurements, we reconstruct the position correctly but overestimate the width.
The resulting NRMSE of 1.71 is much higher than for the synthetic data, which is likely due to the model error of the SWEs.

\subsection{Reconstruction of a Discretized Bathymetry from Synthetic Data}\label{sec:discretized_bathymetry_reconstruction}
As a more realistic and challenging test, we reconstruct the bathymetry in a discretized setting.
Given a spatial discretization $x_i$ of $N = 64$ grid points, we aim to infer the bathymetry height $b_i = b(x_i)$ at each grid point $x_i$.
By not assuming a certain shape of the bathymetry, we increase the dimensionality of the problem significantly.
However, discretizing allows to reconstruct more complex and therefore more realistic bathymetry shapes and requires less prior knowledge.

\paragraph{Impact of the Choice of Prior}
We test two types of prior distributions to enforce different properties on the discretized bathymetry.
First, we use a sparse prior, assuming a mostly flat bathymetry with fairly isolated features.
It is realized by independent Cauchy distributions $\mathcal{C}(\sigma)$ defined by the probability density function
\begin{align}
    \pi_\mathrm{sparse}(\mathbf{b}) = \left(\frac{1}{\pi \sigma}\right)^N \prod_{i=1}^{N} \left(\frac{1}{1 + \left(\frac{b_i}{\sigma}\right)^2}\right)
\end{align}
for each grid point~\cite{kaipioStatisticalInverseProblems2007} where $\sigma$ is the scale parameter controlling sparsity.

The second prior is a smoothness prior, which encodes the assumption that neighboring grid points should have similar bathymetry height.
It is realized using a multivariate Gaussian distribution with a squared exponential covariance matrix~\cite{rameshGATSBIGENERATIVEADVERSARIAL2022}.
Let $i,j$ be indices numbering the equidistant grid points at which the discretized bathymetry is reconstructed.
Then the smoothness prior is given by
\begin{align}
   \mathbf{b} \sim\mathcal{N}(\mathbf{0}, \mathbf{\Sigma}), \quad \Sigma_{i,j}(\sigma^2, l) = \sigma^2 \exp\left(-\frac{(i - j)^2}{l^2}\right),\label{eq:smoothness}
\end{align}
where $\sigma^2$ controls the overall variance of the bathymetry's height and $l\in \mathbb{N}$ is the length scale parameter controlling its smoothness.
We refer to the smoothness prior by $\Sigma(\sigma^2, l)$.

For MCMC sampling, we compute a single chain starting from a flat bathymetry $b_i=0$ for all grid points $i$.
While a broader range of more accurate starting values could likely improve results or at least reduce solution times,  this choice matches the zero bathymetry initial guess used in the PDE-constrained approach against which we compare most closely~\cite{angel24bathy} and therefore seems like a fair comparison.
We experiment with two different proposal distributions.
The first uses independent Gaussian proposals for each grid point with a variance of $\sigma^2=10^{-6}$ while the second is a multivariate Gaussian proposal distribution with a squared exponential covariance matrix like in the smoothness prior~\eqref{eq:smoothness}.
With the second proposal strategy, we obtain similar random steps for neighbouring grid points and avoid proposing bathymetries with large variations between adjacent points, which are unlikely to be accepted.

\begin{figure}[t]
    \includegraphics{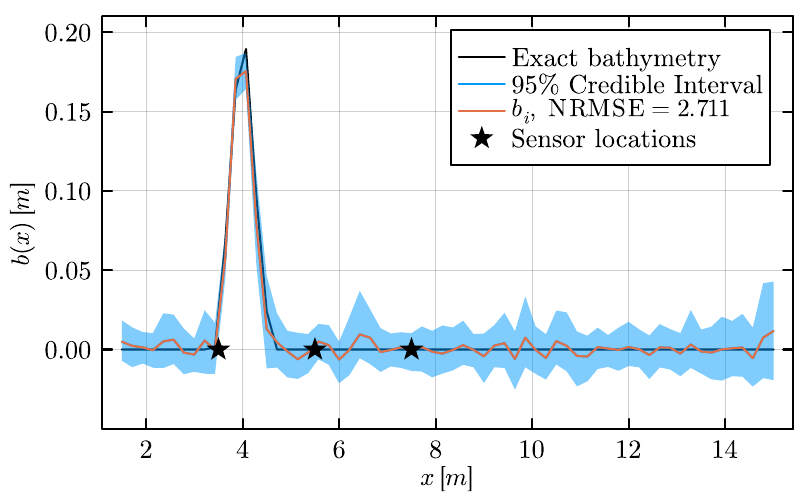}
     \caption{The orange line shows the mean of the MCMC samples after burnin using a combination of the sparse prior with $\sigma=0.01$ and the smoothness prior with ${\sigma^2=0.005}$ and $l=2$.
            We use a multivariate Gaussian proposal distribution with a squared exponential covariance matrix.
            The blue shaded area displays the 95\% credible interval of the MCMC samples.
            The black graph shows the exact bathymetry used to generate the synthetic measurements.
            Note that the difference in scaling of the x- and y-axis.
            The reported relative $L_2$ error is 14.79\% and the relative $\mathrm{L}_\infty$ error is 8.82\%.
     }\label{fig:toy_discretized_result}
\end{figure}

Table~\ref{tab:toy_metrics}  shows the errors in the reconstructed bathymetry for different combinations of prior and proposal distributions.
We use the mean of the 4000 MCMC samples after a burnin of 1000 samples as the reconstructed bathymetry $b_i$ and the exact bathymetry used to generate the synthetic measurements as $b(x_i)$ in~\eqref{eq:nrmse}.
A combination of the sparse and smoothness prior with a smooth proposal distribution leads to the smallest errors and will therefore also be used for reconstruction from experimental data in \S\ref{subsec:experimental}.
The reconstructed bathymetry is shown in Figure~\ref{fig:toy_discretized_result}.
It is reconstructed with low uncertainty around the peak and only small uncertainty in flat areas.
The shape and amplitude of the bathymetry are captured with high accuracy.

Comparing these results to the adjoint optimization approach for synthetic data~\cite{angel24bathy}, we can see that the Bayesian inference approach is able to achieve a more accurate reconstruction of the bathymetry.
Errors improved from a NRMSE ($L_2, L_\infty$) of 10.76 (61.53, 42.75) to a NRMSE ($L_2, L_\infty$) of 2.71 (14.79, 8.82).
The improvement is also reflected in the better reconstruction of the amplitude of the bathymetry with a height of $0.175$ at the peak compared to $0.2$ for the real bathymetry.
With a height of $0.105$ the adjoint approach substantially underestimated the height of the peak.
The Bayesian approach also avoids the spurious hills in the flat area behind the peak that were visible in the solution produced by the PDE-constrained reconstruction.

\begin{table}
        \caption{Error metrics for different combinations of prior distributions and proposal distributions for the discretized bathymetry reconstruction on synthetic data.
        For the sparse prior, we use $\mathcal{C}(0.01)$ and for the smoothness prior we use $\Sigma(0.005, 2)$.
    The NRMSE is calculated using equation \eqref{eq:nrmse}. The values are rounded to two decimal places.}\label{tab:toy_metrics}
    \centering
    \begin{tabular}{l|llccc}
    	\toprule
        Data &Prior & Proposal & NRMSE &$\mathrm{L}_2$ [\%] & $\mathrm{L}_\infty$ [\%]\\
        \midrule
        \multirow{3}{*}{Synthetic}&sparse+smooth &  $\mathcal{N}(0,2\cdot10^{-6})$ & 9.31 & 50.77 & 28.79\\

        &sparse+smooth & $\Sigma(2 \cdot 10^{-6}, 2)$ & 2.71 & 14.79 & 8.82\\

        &sparse & $\Sigma(2 \cdot 10^{-6}, 2)$ & 3.78 & 20.61 & 15.24\\
        \midrule
        \multirow{1}{*}{Experiment}& sparse+smooth & $\Sigma(10^{-6}, 2)$ & 10.70& 58.37 & 35.05\\
        \bottomrule
    \end{tabular}
\end{table}

\subsection{Reconstruction of a Bathymetry from Experimental Data}\label{subsec:experimental}
\begin{figure}
    \includegraphics{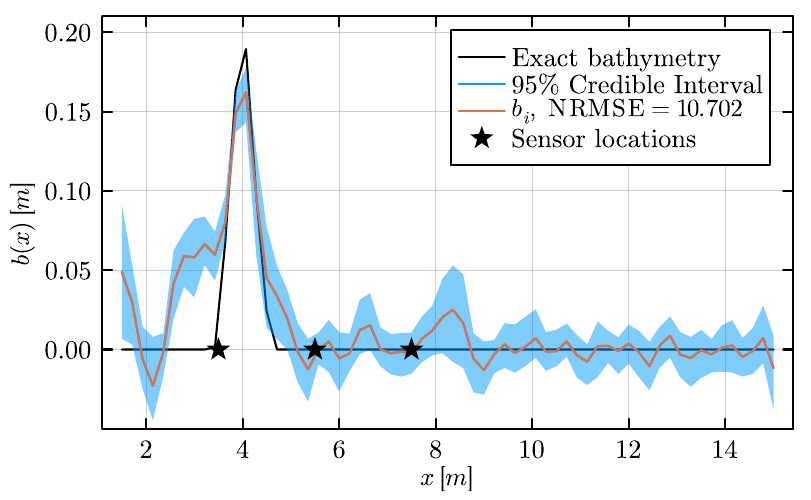}
        \caption{The orange graph shows the mean of the MCMC samples after burnin using the sparse prior with $\sigma=0.01$.
                As proposal we use a multivariate Gaussian proposal distribution with a squared exponential covariance matrix.
                The blue shaded area displays the 95\% credible interval of the MCMC samples.
                The black graph shows the experimental bathymetry.
 	        Note that the difference in scaling of the x- and y-axis.
                The reported NRMSE is 10.70, the relative $L_2$ error is 58.37\% and the relative $\mathrm{L}_\infty$ error is 35.05\%.
        }\label{fig:exp_discretized_result}
\end{figure}

As the final test, we apply the best performing setup from \S\ref{sec:discretized_bathymetry_reconstruction} to the same experimental data used by Angel et al.~\cite{angel24bathy} and compare the performance of Bayesian inversion against their adjoint optimization approach.
The reconstructed bathymetry is shown in Figure~\ref{fig:exp_discretized_result}.
Also from experimental data, the Bayesian approach is able to reconstruct the amplitude of the bathymetry and the peak with low uncertainty.
We achieve a height of $0.16$~m at the peak, slightly lower than with synthetic data, but still closer to the exact height of $0.2$ m than the adjoint optimization approach with $0.111$ m.

The method has some difficulties in reconstructing the shape of the bathymetry before the bathymetry, where spurious peaks are found with high certainty.
Another spurious peak is reconstructed at around $8$~m, but with high uncertainty.
Note that this spurious peak was also present in the adjoint optimization approach.
Tests have shown that further strengthening the sparsity prior can remove the second peak, but at the cost of a less accurate reconstruction of the main peak height and shape.

If the measurements of the sensor at 7.5m are removed from consideration, the second peak also vanishes from the reconstruction.
This indicates that it is likely an artifact of the model error due to dispersive effects neglected in the SWE.
The NRMSE of the reconstruction is 10.70, the relative $\mathrm{L}_2$ error is 58.37 \% and the relative $\mathrm{L}_\infty$ error is 35.05\%.
Comparing those results to the adjoint optimization approach~\cite{angel24bathy}, the Bayesian inference approach is able to achieve a better reconstruction of the bathymetry and reduces the NRMSE ($L_2, L\infty$) by approximately 35\% (38\%, 32\%).

\section{Conclusion}\label{sec:conclusion}
This paper present a Bayesian inference approach for bathymetry reconstruction from time-resolved point-measurements of wave heights.
The goal was to reconstruct a roughly Gaussian-shaped bathymetry in wave flume from discrete sensor measurements over time.

To understand the difficulties of the problem and limitations of the approach, the method is first tested on a low-dimensional parameterized bathymetry with simulated data.
We reconstruct the bathymetry parameterized as a Gaussian defined by only two parameters, width and position.
The results show the method is able to reconstruct the bathymetry in the crucial area around the sensor positions, but struggles if it is positioned too far away from the sensors.
By looking at the posterior distribution, we could highlight the impact of choosing different priors, including the difficulties arising from multiple local extrema in the posterior landscape.

The approach is then tested for the more demanding test case of a discretized bathymetry.
We use synthetic measurements to calibrate the method and to find suitable prior and proposal distributions for MCMC sampling.
The best results were achieved using a combination of a sparse and a smoothness prior and a multivariate Gaussian proposal distribution with a squared exponential covariance matrix.
The calibrated method is then used to reconstruct a real bathymetry from experimental measurements in a wave flume.
Its accuracy is compared to a PDE-constrained optimization approach~\cite{angel24bathy}.

For both the synthetic and experimental data, the Bayesian approach achieved a lower NRMSE with 2.71\% and 10.70\% respectively, compared to 10.76\% and 16.42\% for the adjoint optimization method.
It also produced a better estimate of the bathymetry's height, while the quality of the reconstruction of the overall shape of the bathymetry was similar for both methods.
The Bayesian approach also produces estimates for the uncertainty of the reconstruction, which is not easily possible for the adjoint optimization.

The paper demonstrates the potential usefulness of Bayesian inversion for bathymetry reconstruction.
Future work could focus on integrating a more efficient sampling process, such as Hamiltonian Monte Carlo, No-U-Turn sampler or the Metropolis-adjusted Langevin algorithm, to further improve results.
Instead of MCMC, variational inference methods could be tested to approximate the posterior leveraging machine learning techniques~\cite{rameshGATSBIGENERATIVEADVERSARIAL2022}.

\section*{CRediT authorship contribution statements}
\textbf{Lars Stietz} Writing -- review \& editing, Writing -- original draft, Visualization, Software, Methodology, Formal analysis, Conceptualization.
\textbf{Sebastian G\"otschel} Writing -- review \& editing, Conceptualization, Methodology, Supervision.
\textbf{Peter Schleper} Writing -- review \& editing, Conceptualization, Methodology, Supervision.
\textbf{Daniel Ruprecht} Writing -- review \& editing, Conceptualization, Methodology, Supervision.

\section*{Declaration of Competing Interest}
The authors declare that they have no known competing financial interests or personal relationships that could have appeared to influence the work reported in this paper.

\section*{Data Availability}
The measurement data used in this work was recorded by Angel et al. \cite{dataset} and is publicly available at \url{https://doi.org/10.15480/882.9403} \cite{dataset}.

\section*{Acknowledgements}
We acknowledge DASHH, Data Science in Hamburg - Helmholtz Graduate School for the Structure of Matter, for financial support.

\bibliographystyle{elsarticle-num}
\bibliography{references}

@inproceedings{SeelingerEtAl2021,
author = {Seelinger, Linus and Reinarz, Anne and Rannabauer, Leonhard and Bader, Michael and Bastian, Peter and Scheichl, Robert},
title = {High Performance Uncertainty Quantification with Parallelized Multilevel Markov Chain Monte Carlo},
year = {2021},
doi = {10.1145/3458817.3476150},
booktitle = {Proceedings of the International Conference for High Performance Computing, Networking, Storage and Analysis},
articleno = {75},
numpages = {15},
series = {SC '21}
}

@article{angel24bathy,
title = {Bathymetry Reconstruction from Experimental Data using PDE-Constrained Optimisation},
journal = {Computers \& Fluids},
volume = {278},
pages = {106321},
year = {2024},
doi = {10.1016/j.compfluid.2024.106321},
author = {Judith Angel and Jörn Behrens and Sebastian Götschel and Marten Hollm and Daniel Ruprecht and Robert Seifried}
}

@article{burns20dedalus,
       author = {{Burns}, Keaton J. and {Vasil}, Geoffrey M. and {Oishi}, Jeffrey S. and {Lecoanet}, Daniel and {Brown}, Benjamin P.},
        title = "{Dedalus: A flexible framework for numerical simulations with spectral methods}",
      journal = {Physical Review Research},
         year = 2020,
        month = apr,
       volume = {2},
       number = {2},
        pages = {023068},
          doi = {10.1103/PhysRevResearch.2.023068},
}

@article{hastingsMonteCarloSampling1970,
  title = {Monte {{Carlo}} sampling methods using {{Markov}} chains and their applications},
  author = {Hastings, W. K.},
  year = 1970,
  journal = {Biometrika},
  volume = {57},
  number = {1},
  pages = {97--109},
  doi = {10.1093/biomet/57.1.97},
}

@article{metropolisEquationStateCalculations1953,
  title = {Equation of {{state calculations}} by {{fast computing machines}}},
  author = {Metropolis, Nicholas and Rosenbluth, Arianna W. and Rosenbluth, Marshall N. and Teller, Augusta H. and Teller, Edward},
  year = 1953,
  journal = {The Journal of Chemical Physics},
  volume = {21},
  number = {6},
  pages = {1087--1092},
  doi = {10.1063/1.1699114},
}

@misc{dataset,
	title = {Data artefact: Bathymetry reconstruction from experimental data with PDE-constrained optimisation},
	author = {Judith Angel and Jörn Behrens and Sebastial Götschel and Marten Hollm and Daniel Ruprecht and Robert Seifried},
	doi = {10.15480/882.9403},
	year = {2024}
}

@article{kaipioStatisticalInverseProblems2007,
  title = {Statistical Inverse Problems: {{Discretization}}, Model Reduction and Inverse Crimes},
  author = {Kaipio, Jari and Somersalo, Erkki},
  year = 2007,
  journal = {Journal of Computational and Applied Mathematics},
  series = {Special {{Issue}}: {{Applied Computational Inverse Problems}}},
  volume = {198},
  number = {2},
  pages = {493--504},
  doi = {10.1016/j.cam.2005.09.027},
}

@book{kaipioStatisticalComputationalInverse2005,
  title = {Statistical and Computational Inverse Problems},
  author = {Kaipio, Jari and Somersalo, Erkki},
  year = 2005,
  journal = {Springer e-books},
  series = {Applied Mathematical Sciences},
  number = {v. 160},
  publisher = {Springer},
  doi = {10.1007/b138659}
}

@misc{rameshGATSBIGENERATIVEADVERSARIAL2022,
  title = {{{GATSBI}}: {{Generative Adversarial Training for Simulation-Based Inference}}},
  author = {Ramesh, Poornima and Lueckmann, Jan-Matthis and Boelts, Jan and {Tejero-Cantero}, {\'A}lvaro and Greenberg, David S and Gon{\c c}alves, Pedro J and Macke, Jakob H},
  year = 2022,
  doi = {10.48550/arXiv.2203.06481}
}

@article{robertsOptimalScalingVarious2001a,
  title = {Optimal Scaling for Various {{Metropolis-Hastings}} Algorithms},
  author = {Roberts, Gareth O. and Rosenthal, Jeffrey S.},
  year = 2001,
  journal = {Statistical Science},
  volume = {16},
  number = {4},
  doi = {10.1214/ss/1015346320},
}

@article{hoffmanNoUTurnSamplerAdaptively,
  title={The No-U-Turn sampler: adaptively setting path lengths in Hamiltonian Monte Carlo.},
  author={Hoffman, Matthew D and Gelman, Andrew and others},
  journal={J. Mach. Learn. Res.},
  volume={15},
  number={1},
  pages={1593--1623},
  doi={10.48550/arXiv.1111.4246},
  year={2014}
}

@article{duaneHybridMonteCarlo1987,
  title = {Hybrid {{Monte Carlo}}},
  author = {Duane, Simon and Kennedy, A. D. and Pendleton, Brian J. and Roweth, Duncan},
  year = 1987,
  journal = {Physics Letters B},
  volume = {195},
  number = {2},
  pages = {216--222},
  doi = {10.1016/0370-2693(87)91197-X},
}

@article{robertsExponentialConvergenceLangevin1996,
  title = {Exponential Convergence of {{Langevin}} Distributions and Their Discrete Approximations},
  author = {Roberts, Gareth O. and Tweedie, Richard L.},
  year = 1996,
  journal = {Bernoulli},
  volume = {2},
  number = {4},
  pages = {341--363},
}

@article{cotterMCMCMethodsFunctions2013,
  title = {{{MCMC Methods}} for {{Functions}}: {{Modifying Old Algorithms}} to {{Make Them Faster}}},
  shorttitle = {{{MCMC Methods}} for {{Functions}}},
  author = {Cotter, S. L. and Roberts, G. O. and Stuart, A. M. and White, D.},
  year = 2013,
  journal = {Statistical Science},
  volume = {28},
  number = {3},
  pages = {424--446},
  doi = {10.1214/13-STS421},
}

@misc{lasithadhikariInvertingShoreBathymetry2016,
  title = {Inverting for {{Near Shore Bathymetry}} from {{Surface Wave Properties}}},
  author = {{Lasith Adhikari} and {Charnelle Bland} and {Lopamudra Chakravarty} and {Wenbin Dong} and {Olaniyi Samuel Iyiola} and Muldoon, Gail and Seinen, Clint and Jenkins, Lea and Hesser, Ty and Farthing, Matthew},
  year = 2016,
  doi = {10.13140/RG.2.2.30230.14409},
}

@article{collinsBathymetricInversionUncertainty2020,
  title = {Bathymetric {{Inversion}} and {{Uncertainty Estimation}} from {{Synthetic Surf-Zone Imagery}} with {{Machine Learning}}},
  author = {Collins, Adam and Brodie, Katherine and Bak, Andrew Spicer and Hesser, Tyler and Farthing, Matthew and Lee, Jonghyun and Long, Joseph},
  year = 2020,
  journal = {Remote Sensing},
  volume = {12},
  number = {20},
  pages = {3364},
  doi = {10.3390/rs12203364},
}

@article{stuartInverseProblemsBayesian2010a,
  title = {Inverse Problems: {{A Bayesian}} Perspective},
  author = {Stuart, A. M.},
  year = 2010,
  journal = {Acta Numerica},
  volume = {19},
  pages = {451--559},
  doi = {10.1017/S0962492910000061},
}

@misc{noureddineDirectApproachDetection2025,
  title = {A {{Direct Approach}} for {{Detection}} of {{Bottom Topography}} in {{Shallow Water}}},
  author = {Noureddine, Lamsahel and Rosier, Carole},
  year = 2025,
  number = {arXiv:2510.03505},
  doi = {10.48550/arXiv.2510.03505},
}

@article{heRemoteSensingShallow2024,
  title = {Remote Sensing for Shallow Bathymetry: {{A}} Systematic Review},
  shorttitle = {Remote Sensing for Shallow Bathymetry},
  author = {He, Jinchen and Zhang, Shuhang and Cui, Xiaodong and Feng, Wei},
  year = 2024,
  journal = {Earth-Science Reviews},
  volume = {258},
  pages = {104957},
  doi = {10.1016/j.earscirev.2024.104957},
}

@article{vasanInverseWaterWave2013,
  title = {The Inverse Water Wave Problem of Bathymetry Detection},
  author = {Vasan, Vishal and Deconinck, Bernard},
  year = 2013,
  journal = {Journal of Fluid Mechanics},
  volume = {714},
  pages = {562--590},
  doi = {10.1017/jfm.2012.497},
}

@article{oadesEvaluationNearshoreBathymetric2023,
  title = {Evaluation of Nearshore Bathymetric Inversion Algorithms Using Camera Observations and Synthetic Numerical Input of Surface Waves during Storms},
  author = {Oades, Elora M. and Mulligan, R. P. and Palmsten, M. L.},
  year = 2023,
  journal = {Coastal Engineering},
  volume = {184},
  pages = {104338},
  doi = {10.1016/j.coastaleng.2023.104338},
}

@article{sellierInverseProblemsFree2016,
  title = {Inverse Problems in Free Surface Flows: A Review},
  author = {Sellier, Mathieu},
  year = 2016,
  journal = {Acta Mechanica},
  volume = {227},
  number = {3},
  pages = {913--935},
  doi = {10.1007/s00707-015-1477-1},
}

@article{wuAdjointbasedHighorderSpectral2023a,
  title = {Adjoint-Based High-Order Spectral Method of Wave Simulation for Coastal Bathymetry Reconstruction},
  author = {Wu, Jie and Hao, Xuanting and Li, Tianyi and Shen, Lian},
  year = 2023,
  journal = {Journal of Fluid Mechanics},
  volume = {972},
  pages = {A41},
  doi = {10.1017/jfm.2023.733},
}

@article{Bhagavan2024,
  doi = {10.21105/joss.06917},
  year = {2024},
  publisher = {The Open Journal},
  volume = {9},
  number = {101},
  pages = {6917},
  author = {Sathvik Bhagavan and Bart de Koning and Shubham Maddhashiya and Christopher Rackauckas},
  title = {DataInterpolations.jl: Fast Interpolations of 1D data},
  journal = {Journal of Open Source Software}
}

@article{10.1145/3711897,
  author = {Fjelde, Tor Erlend and Xu, Kai and Widmann, David and Tarek, Mohamed and Pfiffer, Cameron and Trapp, Martin and Axen, Seth D. and Sun, Xianda and Hauru, Markus and Yong, Penelope and Tebbutt, Will and Ghahramani, Zoubin and Ge, Hong},
  title = {Turing.jl: A General-Purpose Probabilistic Programming Language},
  year = {2025},
  doi = {10.1145/3711897},
  journal = {ACM Trans. Probab. Mach. Learn.},
}

@InProceedings{pmlr-v84-ge18b,
  title = {Turing: A Language for Flexible Probabilistic Inference},
  author = {Ge, Hong and Xu, Kai and Ghahramani, Zoubin},
  booktitle = {Proceedings of the Twenty-First International Conference on Artificial Intelligence and Statistics},
  pages = {1682--1690},
  year = {2018},
  volume = {84},
  series = {Proceedings of Machine Learning Research},
  month = {09--11 Apr},
  url = {https://proceedings.mlr.press/v84/ge18b.html},
}

@article{behrens2010new,
  title={A New Multi-Sensor Approach to Simulation Assisted Tsunami Early Warning},
  author={Behrens, J{\"o}rn and Androsov, Alexey and Babeyko, AY and Harig, Sven and Klaschka, Florian and Mentrup, Lars},
  journal={Natural Hazards and Earth System Sciences},
  volume={10},
  number={6},
  pages={1085--1100},
  year={2010},
  doi = {10.5194/nhess-10-1085-2010}
}

@article{dalrymple2011rip,
  title={Rip currents},
  author={Dalrymple, Robert A and MacMahan, Jamie H and Reniers, Ad JHM and Nelko, Varjola},
  journal={Annual Review of Fluid Mechanics},
  volume={43},
  number={1},
  pages={551},
  year={2011},
  doi={10.1146/annurev-fluid-122109-160733}
}

@article{neal2021estimating,
  title={Estimating River Channel Bathymetry in Large Scale Flood Inundation Models},
  author={Neal, Jeffrey and Hawker, Laurence and Savage, James and Durand, Michael and Bates, Paul and Sampson, Christopher},
  journal={Water Resources Research},
  volume={57},
  number={5},
  pages={e2020WR028301},
  year={2021},
  doi={10.1029/2020WR028301}
}

@article{ruppenthal2026bathymetry,
  title={Bathymetry Reconstruction via Optimal Control in Well-Balanced Finite Element Methods for the Shallow Water Equations},
  author={Ruppenthal, Falko and Kuzmin, Dmitri},
  journal={arXiv preprint arXiv:2603.11813},
  year={2026},
  doi={10.48550/arXiv.2603.11813}
}

@software{myCode,
  author       = {Stietz, Lars Olaf and
                  Ruprecht, Daniel},
  title        = {Code: Bathymetry Reconstruction by Bayesian
                   Inference
                  },
  month        = apr,
  year         = 2026,
  publisher    = {Zenodo},
  doi          = {10.5281/zenodo.19472936},
  }

\end{document}